\documentclass{camera-hepex}
\usepackage{epsfig}
\begin{document}

\title{MEASUREMENTS OF\\SINGLE W PRODUCTION AT LEP2}
\author{Andrea Valassi}
\organization{CERN, EP Division, 1211 Geneva 23, Switzerland}
\maketitle

\newcommand{\etal}       {{\it et al.}}
\newcommand{\PLB}[3]     {Phys.\ Lett.\ {\bf B#1} (#2) #3}
\newcommand{\PTP}[3]     {Prog.\ Theor.\ Phys.\ \textbf{#1} (#2) #3}
\newcommand{\NPB}[3]     {Nucl.\ Phys.\ {\bf B#1} (#2) #3}
\newcommand{\CPC}[3]     {Comp.\ Phys.\ Comm.\ {\bf #1} (#2) #3}

\newcommand{\Aleph}      {\mbox{A{\sc leph}}}
\newcommand{\Delphi}     {\mbox{D{\sc elphi}}}
\newcommand{\Ltre}       {\mbox{L{\sc 3}}}
\newcommand{\Opal}       {\mbox{O{\sc pal}}}
\newcommand{\Conf}       {\mbox{C{\sc onf}}}
\newcommand{\Lep}        {\mbox{L{\sc ep}}}
\newcommand{\Lept}       {\mbox{L{\sc ep2}}}
\newcommand{\Lepewwg}    {\mbox{L{\sc epewwg}}}
\newcommand{\Xsec}       {\mbox{X{\sc sec}}}

\newcommand{\Grace}      {\mbox{\tt grc4f}}
\newcommand{\WTO}        {\mbox{WTO}}
\newcommand{\WPHACT}     {\mbox{WPHACT}}

\def\qq      { {\mathrm{q\bar{q}}} }
\def\e       { {\mathrm{e}} }
\def\f       { {\mathrm{f}} }
\def\W       { {\mathrm{W}} }
\def\Z       { {\mathrm{Z}} }
\def\WWg     { {\mathrm{WW}\gamma} }
\def\WWZ     { {\mathrm{WWZ}} }
\def\Dgiz    { \Delta\mathrm{g}_1^\Z }
\def\Dkg     { \Delta\kappa_\gamma }
\def\lg      { \lambda_\gamma }

\vspace*{-0.4cm}
In the energy region of \Lept\ and above, 
four fermion final states in $\e^+\e^-$ collisions can 
be produced by Feynman diagrams involving two, one, or zero resonant bosons.
The four fermion process $\e^+\e^-\rightarrow\e\nu\f\bar\f'$ 
is an interesting reaction because it involves \mbox{$t$-channel} diagrams,
whose contribution is enhanced 
when the electron is scattered at very low angles 
and escapes experimental detection along the beam pipe.
Under these conditions,
the process is commonly referred to as single W production, 
$\e^+\e^-\rightarrow\e\nu\W$, 
as it is dominated by diagrams involving one single resonant heavy W boson.

The interest of the experimental study 
of single W production at \Lept\ is twofold. 
First, this process is a powerful probe of possible physics beyond 
the Standard Model, thanks to its strong sensitivity to  
anomalous triple gauge boson couplings~(TGCs)~\cite{bib:tsukamoto}. 
Secondly, the experimental signature of single W events,
with large missing energy and transverse momentum
due to the undetected electron and neutrino,
is also characteristic in the searches for new particles,
and its measurement provides a useful check of the validity 
of the background estimation for these analyses.

\subsection*{Measurements of single W production cross sections}
Single W cross sections at \Lept\ have 
been measured by the four \Lep\ experiments
at energies between 130 and 207 GeV~\cite{bib:swmor01},
in both the hadronic and leptonic decay channels of the W boson.
Hadronic single W events, 
characterised by two acoplanar jets of high invariant mass
accompanied by large transverse missing energy,
are typically selected with 40\% efficiency and 40\% purity,
with large $\W\W\rightarrow\qq\tau\nu$ 
and $\Z\Z\rightarrow\qq\nu\bar{\nu}$ backgrounds.
Leptonic single W events,
with a single isolated electron or muon
(or thin jet for $\tau$ decays) of high $p_T$,
are typically selected with 60\% efficiency and 60\% purity,
the largest backgrounds being 
$\Z\e\e$ or radiative Z events with $\Z\rightarrow\nu\bar{\nu}$,
as well as radiative Bhabha events.
At 189~GeV, 60 (30) events per experiment are typically accepted 
by the hadronic (leptonic) analyses.
Both measurements are limited by statistical errors.

\begin{figure}[t]
  \begin{center}
    \vspace*{-0.3cm}
    \mbox{
      \hspace*{-1cm}
      \epsfig{figure=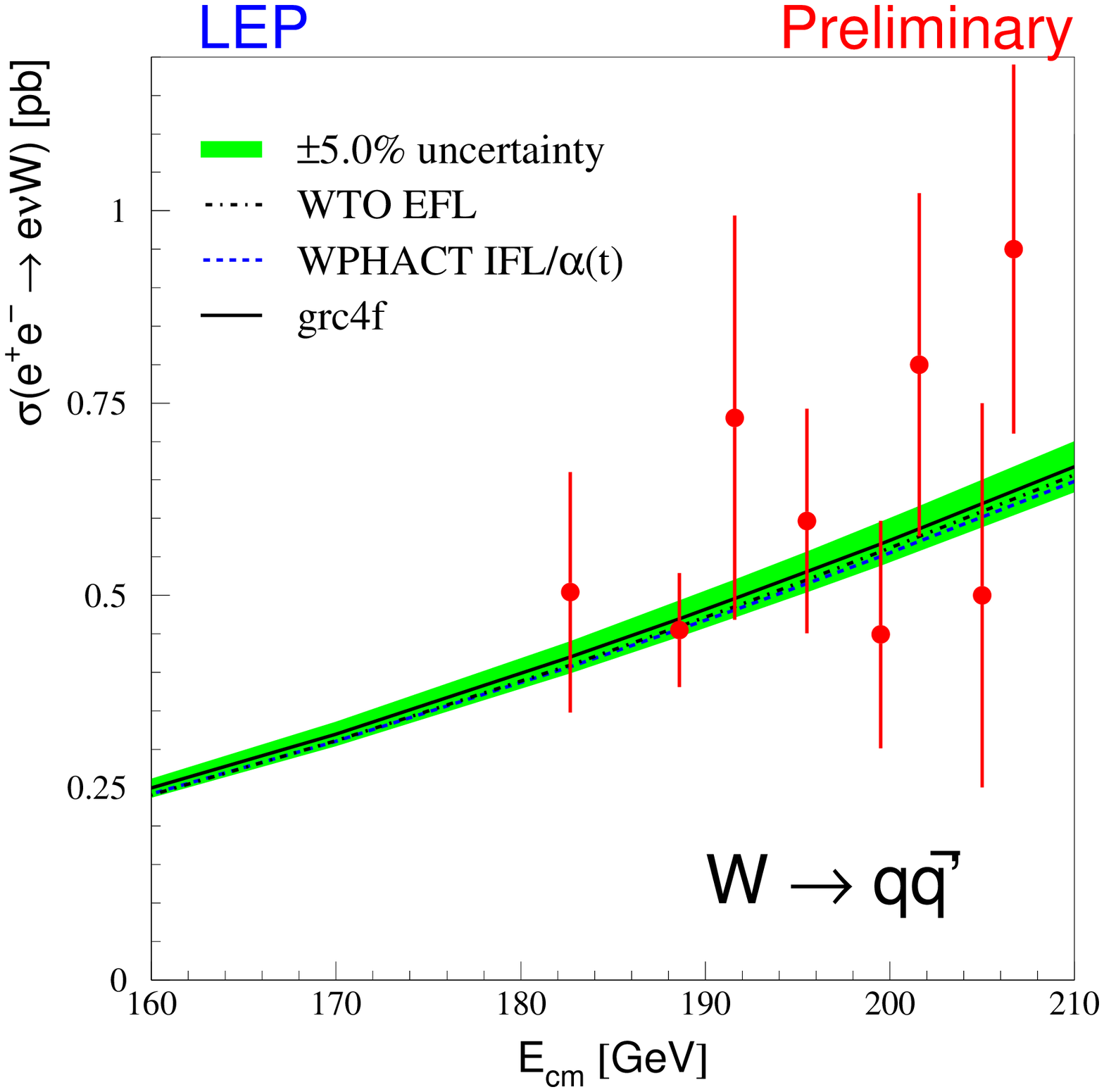,width=8.5cm}
      \hspace*{0.1cm}
      \epsfig{figure=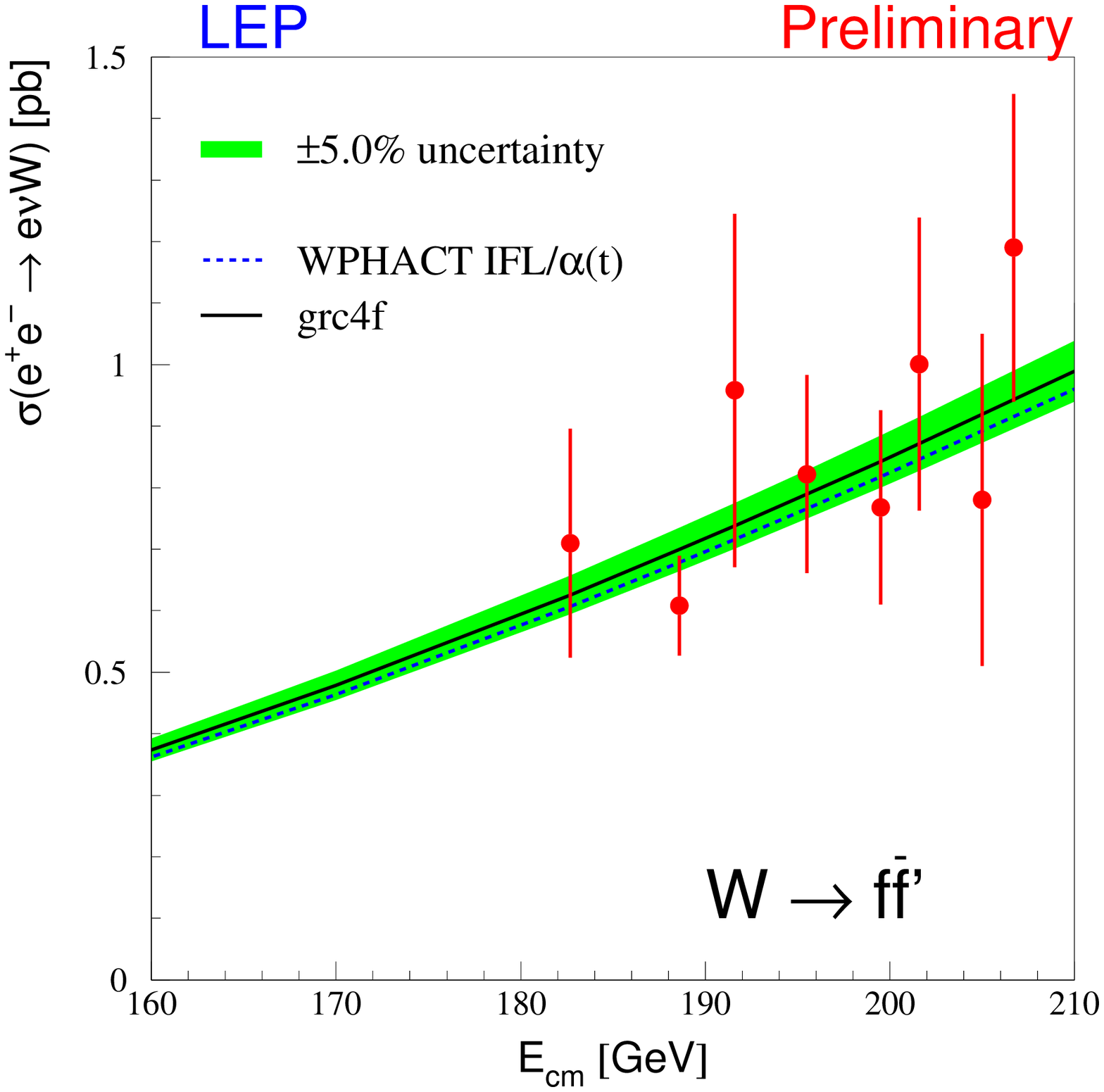,width=8.5cm}
      }
    \vspace*{-1.0cm}
    \caption{
      \Lep\ measurements of the hadronic (left) and total (right) 
      single W production cross section, compared to the predictions 
      of \WTO, \WPHACT\ and \Grace~\cite{bib:mcs}.
      The shaded areas represent the $\pm5$\% uncertainty 
      on the predictions~\cite{bib:accomando}.
      }
    \vspace*{-0.6cm}
    \label{fig:swen_xsec}
  \end{center}
\end{figure}

The combined values of the single W cross sections 
measured by the four \Lep\ experiments 
between 183 and 207 GeV~\cite{bib:swmor01}
are compared in Figure~\ref{fig:swen_xsec}
to several theoretical predictions~\cite{bib:mcs,bib:accomando}.
No deviation of the measurements
from the expectations is observed.
The values in the figures represent single W cross sections 
according to the common \Lep\ definition~\cite{bib:swmor01}, 
where single W production is considered
as the complete $t$-channel subset of Feynman diagrams
contributing to e$\nu_\mathrm{e}$f$\bar{\mathrm{f}}'$ final states,
with additional kinematic cuts
to exclude the regions of phase space 
dominated by multiperipheral diagrams.

\subsection*{Sensitivity to triple gauge boson couplings}
Triple gauge boson couplings, usually described in terms of the 
three parameters $\Dgiz$, $\Dkg$ and $\lg$~\cite{bib:lep2tgcyr},
are measured at \Lept~\cite{bib:terranova} primarily from the analysis 
of W pair events, $\e^+\e^-\rightarrow\W^+\W^-$.
Since this process is sensitive to both the $\WWZ$ and the $\WWg$ vertices,
it is difficult to disentangle the two effects and large correlations 
exist between the fitted values of the couplings from these analyses.
Single W production at \Lept\ is especially interesting 
because it is sensitive to the $\WWg$ vertex alone~\cite{bib:tsukamoto},
the contribution of the WWZ vertex being suppressed 
by the $t$-channel exchange of a Z,
thus providing useful constraints complementary 
to those from W pair events.

\begin{figure}[t]
  \begin{center}
    \vspace*{-1.8cm}
    \mbox{
      \hspace*{-1.0cm}
      \epsfig{figure=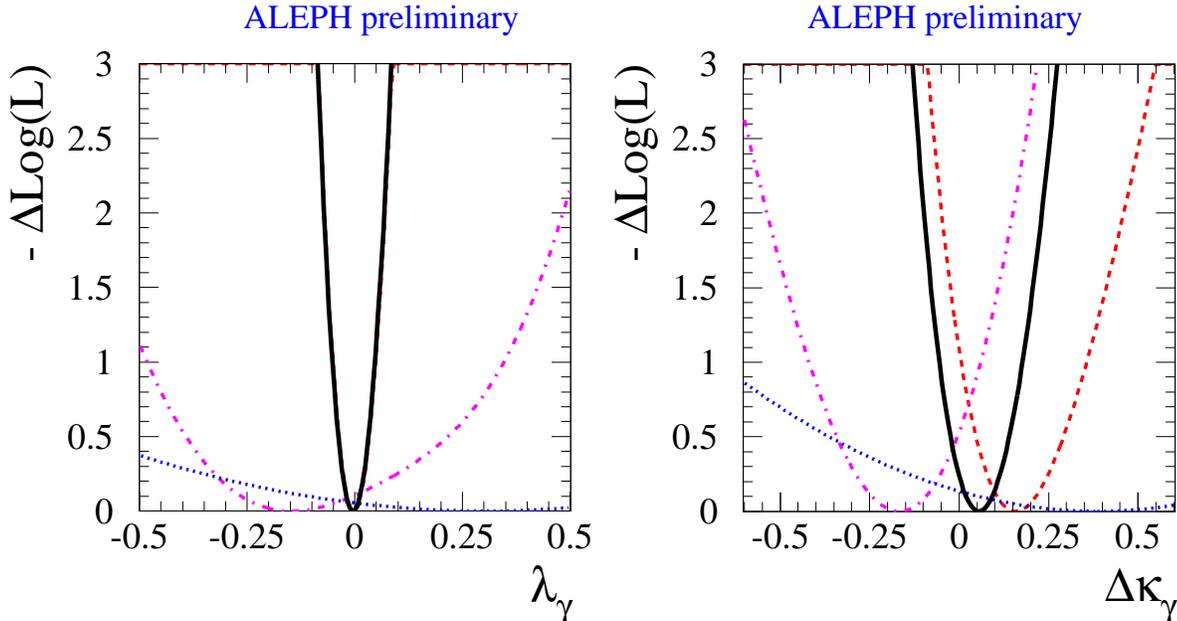,width=17.0cm}
      }
    \vspace*{-8.9cm}
    \caption{
      \Aleph\ preliminary measurements of $\lg$ (left) and $\Dkg$ (right)
      at energies up to 202 GeV~\cite{bib:aletgc202},
      from W pair (dashed red), single $\gamma$ (dotted blue)
      and single W events (dashed-dotted purple) 
      and their combination (solid black).
      }
    \vspace*{-0.4cm}
    \label{fig:tgcs}
  \end{center}
\end{figure}

Measurements of $\Dkg$ and $\lg$ from the analysis of single W events
have been performed
by \Aleph\ at 161--202, \Delphi\ at 189--202, \Ltre\ at 130--202 
and \Opal\ at 189~GeV~\cite{bib:adloswtgc}.
TGCs are extracted via maximum likelihood fits 
to total event rates and differential distributions,
most of the sensitivity coming from event rates.
As expected~\cite{bib:tsukamoto},
for a given integrated luminosity,
the sensitivity of single W production to $\Dkg$
is comparable to that obtained from W pair events,
in spite of the much lower cross section.
This can be seen, for instance, in Figure~\ref{fig:tgcs},
where the \Aleph\ measurements of $\lg$ and $\Dkg$
at energies up to 202 GeV~\cite{bib:aletgc202}
from W pair, single W and single $\gamma$ events are compared and combined.
As the figure also shows, W pair events have a considerably
higher sensitivity to $\lg$ compared to single W events.
Combined \Lep\ results for TGCs from W pair, single $\gamma$ 
and single W events have been presented in Ref.~\cite{bib:terranova}.

\subsection*{Conclusions and acknowledgements}
Single W production has been studied by the four \Lep\ experiments
at all \Lept\ energies between 130 and 207 GeV. 
The process is sensitive to anomalous triple gauge boson couplings,
especially to the $\WWg$ coupling $\Dkg$.
No deviation of the measured cross sections from the theoretical predictions 
of the Standard Model has been observed.

I would like to thank my colleagues from the \Lep\ experiments 
and the \Lep\ WW Working Groups for many useful discussions.
I also warmly thank the organisers of this conference 
for their kind hospitality.

\renewcommand\refname

\end{document}